\documentclass[apj]{emulateapj}

\received{2008 August 13}
\accepted{2008 September 30}

\def\simgt{\lower.5ex\hbox{$\; \buildrel > \over \sim \;$}}
\def\simlt{\lower.5ex\hbox{$\; \buildrel < \over \sim \;$}}

\def\etal{{et~al.}}
\def\amin{\ifmmode^{\prime}\else$^{\prime}$\fi}
\def\asec{\ifmmode^{\prime\prime}\else$^{\prime\prime}$\fi}

\def\simgt{\lower.5ex\hbox{$\; \buildrel > \over \sim \;$}}
\def\simlt{\lower.5ex\hbox{$\; \buildrel < \over \sim \;$}}

\newcommand\asca{{\it ASCA\/}}
\newcommand\chandra{{\it Chandra}}

\newcommand\agile{{\it AGILE\/}}
\newcommand\cgro{{\it CGRO\/}}

\newcommand\pwna{G75.2+0.1}
\newcommand\pwnb{G74.9+1.2 = CTB 87}
\newcommand\mgro{MGRO~J2019+37}

\newcommand\egra{3EG~J2021+3716}
\newcommand\egrb{3EG~J2016+3657}
\newcommand\gev{GeV~J2020+3658}
\newcommand\blazar{B2013+370}

\newcommand\psr{PSR~J2021+3651}
\newcommand\agl{AGL~J2020.5+3653}

\slugcomment{The Astrophysical Journal Letters}

\shorttitle{Discovery of Gamma-Ray Pulsations from PSR J2021+3651}
\shortauthors{Halpern et al.}

\begin{document}

\title{Discovery of High-Energy Gamma-Ray Pulsations from PSR J2021+3651 with AGILE} 

\author{J. P. Halpern\altaffilmark{1}, F.~Camilo\altaffilmark{1},
A.~Giuliani\altaffilmark{2},
E.~V.~Gotthelf\altaffilmark{1}, M.~A. McLaughlin\altaffilmark{3},
R.~Mukherjee\altaffilmark{4},
A.~Pellizzoni\altaffilmark{2}, S.~M. Ransom\altaffilmark{5},
M.~S.~E. Roberts\altaffilmark{6},
and M.~Tavani\altaffilmark{7,8}}

\altaffiltext{1}{Columbia Astrophysics Laboratory, Columbia University,
New York, NY 10027.}

\altaffiltext{2}{INAF/IASF-Milano, I-20133 Milano, Italy.}

\altaffiltext{3}{Department of Physics, West Virginia University,
Morgantown, WV 26501.}

\altaffiltext{4}{Department of Physics and Astronomy, Barnard College,
Columbia University, New York, NY 10027.}

\altaffiltext{5}{National Radio Astronomy Observatory,
Charlottesville, VA 22903.}

\altaffiltext{6}{Eureka Scientific, Inc., Oakland, CA 94602.}

\altaffiltext{7}{INAF/IASF-Roma, I-00133 Roma, Italy.}

\altaffiltext{8}{Dipartimento di Fisica, Universit\`a ``Tor Vergata'', 
I-00133 Roma, Italy.}

\begin{abstract}

Discovered after the end of the {\it Compton Gamma-Ray Observatory}
mission, the radio pulsar \psr\ was long considered a likely
counterpart of the high-energy $\gamma$-ray source 2CG~075+00 = \egra\ = \gev,
but it could not be confirmed due to the lack of a contemporaneous
radio pulsar ephemeris to fold the sparse, archival $\gamma$-ray photons.
Here, we report the discovery of $\gamma$-ray pulsations from \psr\
in the 100--1500 MeV range
using data from the \agile\ satellite gathered over 8 months, folded
on a densely sampled, contemporaneous radio ephemeris obtained for this
purpose at the Green Bank Telescope.  The $\gamma$-ray pulse consists
of two sharp peaks separated by $0.47\pm 0.01$ cycles.  The single radio
pulse leads the first $\gamma$-ray peak by $0.165\pm 0.010$ cycles.
These properties are similar to those of other $\gamma$-ray pulsars,
and the phase relationship of the peaks can be interpreted in the
context of the outer-gap accelerator model for $\gamma$-ray emission.
Pulse-phase resolved images show that there is only one dominant source,
\agl\ = \psr, in the region previously containing confused sources
\egra\ and \egrb.

\end{abstract}

\section{Introduction}

Rotation-powered pulsars are the dominant class of high-energy
($> 100$ MeV) $\gamma$-ray sources in the Galaxy.  The
Energetic Gamma-Ray Experiment Telescope (EGRET) on the
{\it Compton Gamma-Ray Observatory} (\cgro), 
which operated between 1991 and 2000, detected six pulsars,
and three others with less confidence.
(See \citealt{tho04} for a review of these EGRET results.)
It is widely assumed that a large
number of unidentified $\gamma$-ray sources in the Galactic
plane are also pulsars.  Some of these are positionally coincident
with radio pulsars, but do not have enough photons to recover the
pulse period, while others may be radio quiet.  An ephemeris
from radio or X-ray observations is generally
needed to fold the sparse $\gamma$-ray source photons over a
span of months to years in order to build up a statistically
significant pulsed
light curve.  In several cases, a radio pulsar candidate
was only discovered after the end of the \cgro\ mission, so that
the needed contemporaneous ephemeris does not exist.  The typical
``timing noise'' of young pulsars causes a phase ephemeris to
fail when extrapolated over months to years.

Some of the still-unidentified Galactic $\gamma$-ray sources 
date from the 1975 {\it COS B} mission \citep{swa81}.
The target of this work is the {\it COS B} source 2CG 075+00
located in Cygnus.
EGRET possibly resolved 2CG 075+00
into two sources, \egra\ and the weaker \egrb\ \citep{har99}.
Multiwavelength study of these regions by \citet{muk00}
and \citet{hal01} uncovered possible counterparts for \egrb,
but no obvious candidate for \egra.  A smaller error region
was derived by \citet{lam97} by restricting analysis
to energies above 1~GeV, yielding a source \gev\
that overlapped with \egra.  An \asca\ observation
of \gev\ found two compact X-ray sources \citep{rob01},
one of which was discovered to contain a new, energetic
radio pulsar, \psr\ \citep{rob02}.  Its spin period $P = 0.103$~s,
characteristic age $\tau_c \equiv P/2\dot P$ = 17~kyr, and spin-down
luminosity $\dot E = 3.4 \times 10^{36}$ ergs s$^{-1}$ are in the
range of other young pulsars that were detected by EGRET, but its
initial distance estimate of $\sim 12$~kpc implied an unusually
large $\gamma$-ray efficiency (see \S 3.3) if \psr\ is
responsible for \egra.  It was not possible to extrapolate
the radio ephemeris backward to search
for the pulsar in EGRET $\gamma$-ray photons.
By searching in a range around the expected $P$ and $\dot P$,
\citet{mc04} possibly detected pulsations, but only in
two out of eight EGRET viewing periods.
Thus, the proposed identification of \psr\ with \egra\ and
\gev\ could not be confirmed, until now.

The \agile\ ({\it Astro-rivelatore Gamma a Immagini LEggero})
mission of the Italian Space Agency (ASI) was launched on 2007 April 23
into a low-Earth orbit \citep{tav08}.  \agile\ will allow a number of
new $\gamma$-ray pulsars to be found by folding on contemporaneous
pulsar ephemerides that are being constructed by our group
and others using several radio telescopes.  For the
analysis of \psr, we have been using the NRAO Green Bank Telescope
(GBT).  In this Letter, we apply the ephemeris of \psr\
to data obtained by \agile\ during the first year
of the mission.

\section{Observations and Analysis}

The \agile\ GRID (Gamma-Ray Imaging Detector) comprises 
a Silicon-Tungsten tracker \citep{bar01,pre03}
and a CsI(Tl) mini-calorimeter \citep{lab06}.  These provide effective
area of $\sim 500$~cm$^2$ in the 30 MeV--50 GeV range over
a $\sim 2.5$ sr field of view.  We used
GRID observations of the Cygnus region during five
periods between 2007 November 2 and 2008 June 30
that sum to 103 days of observation when
\psr\ was within $40^{\circ}$ of the pointing direction.
GRID photons are time-tagged with a precision of $\sim 1\ \mu$s,
and the arrival times were transformed to the solar system barycenter
in Barycentric Dynamical Time using GPS positioning of the
spacecraft and the \chandra\ measured position of the pulsar,
(J2000.0) R.A. = $20^{\rm h}21^{\rm m}05.46^{\rm s}$,
decl. = $+36^{\circ}51^{\prime}04.8^{\prime\prime}$, uncertainty
$\sim 0.5^{\prime\prime}$ \citep{hes04}.
The accuracy of this procedure is $\sim 10\ \mu$s, and the absolute
times were verified to a precision of $\sim 400\ \mu$s using observations
of the Crab pulsar and others \citep{pel08}.

A phase-connected rotational ephemeris for \psr\ contemporaneous with all
of the \agile\ observing periods was derived from 36 observations obtained
at the GBT between 2007 September 28 and 2008 June 30.  The pulsar was
observed with the GBT Pulsar Spigot \citep{kap05},
yielding total power samples every
$81.92\ \mu$s in each of 768 frequency channels spanning a bandwidth of
600\,MHz centered on 1950\ MHz.  Each observation usually lasted for 5
minutes, from which we derived a time of arrival with typical uncertainty
of 0.2\,ms.  We used the
TEMPO\footnote{See http://www.atnf.csiro.au/research/pulsar/tempo.}
timing software to obtain the
ephemeris using standard methods.  Like many young pulsars, \psr\ exhibits
significant rotational instability.  We parameterized this by fitting for
the rotational frequency and its first six derivatives.  This timing
solution had small unmodeled residual features ($\chi^2_\nu = 2.5$),
with a post-fit rms of 0.33\ ms and maximum of 0.9\ ms.  This precision
is comparable to the absolute accuracy of the \agile\ timing, and ensures
that the effective time resolution of the folded $\gamma$-ray light curve
is $\simlt 0.01\,P$.  The observed radio dispersion measure ($\mbox{DM}
= 369.3\pm0.3$\ pc\ cm$^{-3}$, determined with the addition of a few
observations centered at 1550\ MHz) was used to correct the times
of arrival of the radio pulse to infinite frequency (with resulting
$1\sigma$ uncertainty of 0.6\ ms) for absolute phase comparison with
the $\gamma$-ray pulse.

In choosing the sky region around the pulsar position for extraction
of $\gamma$-ray photons, consideration of the point-spread
function (PSF) of the instrument, which increases with decreasing
energy, must be weighed against the high diffuse $\gamma$-ray
background in this region, which also increases with decreasing
energy, as well as possible contamination from nearby
point sources.
We find that the
pulsed power is maximized if we use an extraction region that is
comparable to the $68\%$ containment radius, which is
$3.5^{\circ}$ at 100~MeV and $1.2^{\circ}$ at 400~MeV \citep{tav08}.
We used $r = 1.2^{\circ}$ for $E \ge 400$ MeV,
and $r = 1.2^{\circ}(400/E)^{1/2}$ for $E < 400$ MeV.
In addition. 
we use only the ``G'' class events, which are identified with good
confidence as photons, and exclude the ``L'' class which suffer
much higher particle background contamination.  This selection
limits the effective area to $\sim 250$~cm$^2$ at 100~MeV.

\begin{figure}[t]
\centerline{
\includegraphics[height=1.15\linewidth,clip=true]{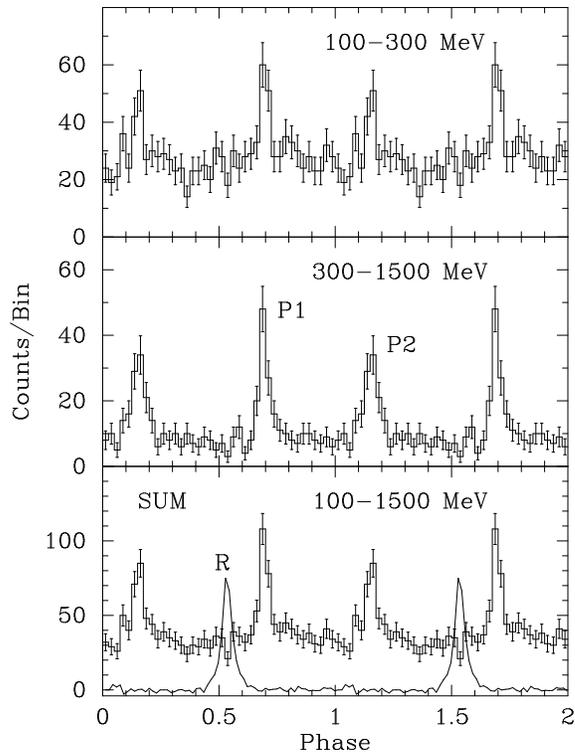}
}
\caption{Folded $\gamma$-ray light curves ({\it histograms})
of \psr\ in two energy bands, and their sum.
Two cycles are plotted, with 40 bins per cycle.
The solid line in the bottom panel is the mean radio pulse at
1950 MHz from the GBT. Phase lags are P2--P1=$0.47\pm 0.01$
cycles and P1--R=$0.165\pm 0.010$ cycles.}
\label{fig:pulse}
\end{figure}

\section{Results and Interpretation}

\subsection{Pulsed Light Curve}

Figure~\ref{fig:pulse} shows the resulting folded light curve
of \psr\ in
two energy bands, $100-300$~MeV and $300-1500$~MeV.
The pulse shape consists of two sharp peaks,
labeled P1 and P2, with 
separation of $0.47 \pm 0.01$ cycles \cite[consistent with the separation
in the tentative detections reported in EGRET data by][]{mc04}.
This structure
is similar to that seen with much better signal-to-noise ratio in
other pulsars.  For comparison, peak separations are 0.38,
0.43, and 0.49 for the Crab, Vela, and Geminga, respectively.
As in the other $\gamma$-ray pulsars, the
radio pulse of \psr\ leads the $\gamma$-ray P1 by a 
significant phase, in this case $0.165 \pm 0.010$ cycles.
In Vela, the radio leads by 0.13 cycles.
The Crab pulsar has multiple broad radio peaks that
continue to high frequency, and coincide with
higher-energy emission from infrared through X-ray.  However,
if we identify the narrow low-frequency component
\citep[LFC:][]{mof96,mof99} as the usual polar-cap core emission analogous
to other pulsars, the $\gamma$-ray--radio lag is 0.11 cycles.  Alternatively,
the Crab's even lower frequency ``precursor'' pulse, which leads
the $\gamma$-ray P1 by 0.056 cycles, may be the core component.
Even in the case of PSR~B1706$-$44, where \agile\ resolved
two peaks separated by 0.24 cycles from the dominant ``bridge''
emission, the radio pulse leads P1 by 0.21 cycles \citep{pel08}.

\begin{figure}[t]
\centerline{
\includegraphics[height=0.85\linewidth,clip=true]{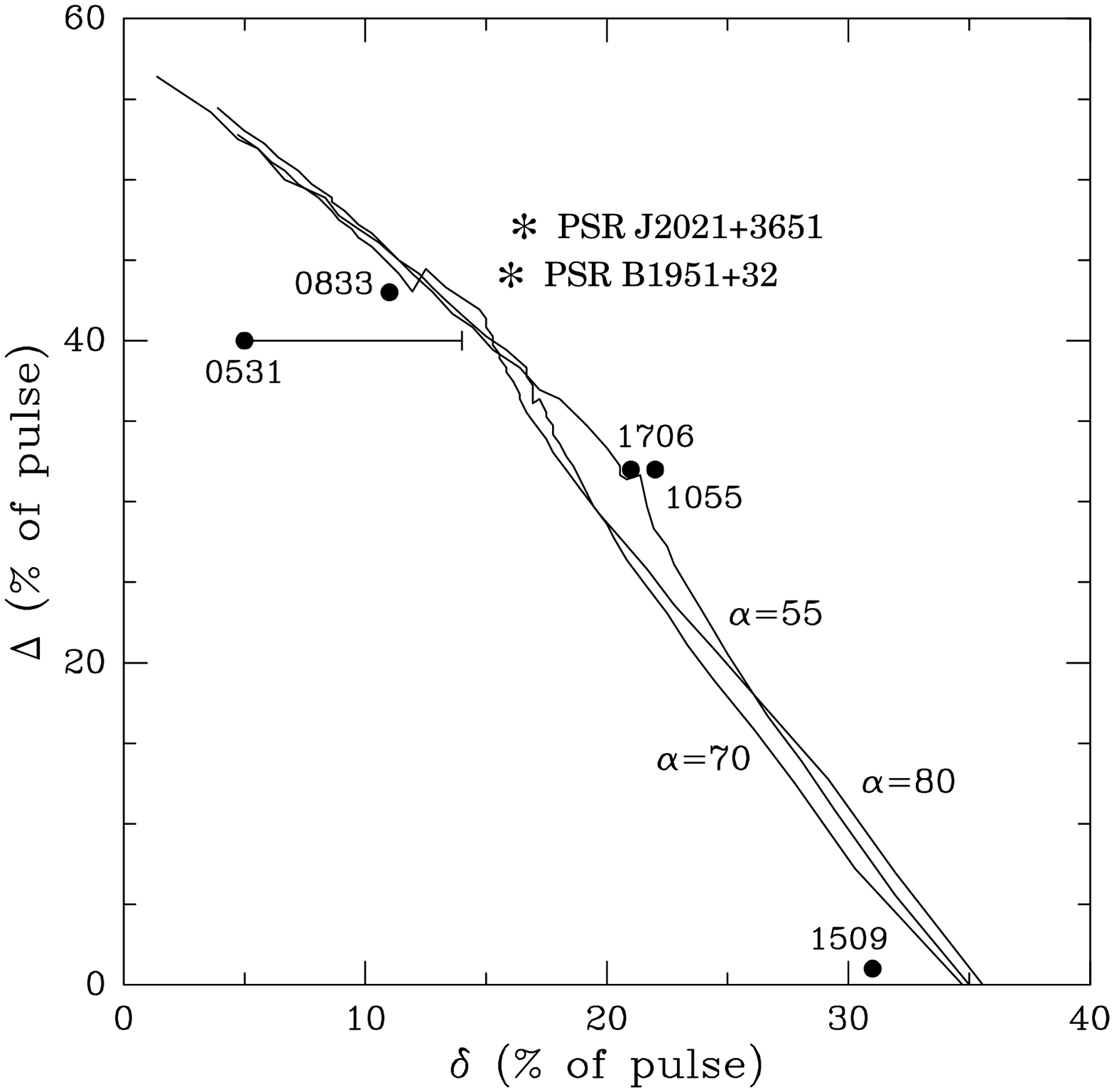}
}
\caption{Reproduction of Figure~3 of \citet{rom95} comparing
$\gamma$-ray peak
separation $\Delta$ with $\gamma$-ray--radio lag $\delta$
for \cgro\  pulsars ({\it filled circles}).  The $\delta$
of the Crab pulsar (0531) is uncertain due to the
ambiguous identification of its core radio pulse.
The location of \psr\ is marked with an asterisk,
as is the closely situated PSR~B1951+32 \citep{ram05}.
Model predictions ({\it solid lines}) for different magnetic axis
inclination angles $\alpha$ are similar, while viewing angle $\zeta$
increases along the curves from lower right to upper left.
}
\label{fig:delta}
\end{figure}

The outer-gap model of $\gamma$-ray pulse shapes of
\citet{rom95} predicts an inverse dependence of the
$\gamma$-ray peak separation on the $\gamma$-ray--radio
lag, with the principal controlling parameter being viewing
angle $\zeta$ from the rotation axis.
Large $\zeta$ results in large peak separation,
and small $\gamma$-ray--radio lag.
\psr\ is roughly in accord with this prediction
(Fig.~\ref{fig:delta}).  In this picture,
pulsars that have peak separations close to 0.5, such as
Vela, Geminga, and \psr, have $\zeta$ close to $90^{\circ}$.
Such an angle for \psr\ is also supported by the
modeling of its X-ray pulsar wind nebula (PWN) \pwna\
\citep{hes04,van08}, which can be understood as a nearly
edge-on torus,
indicating a pulsar rotation axis nearly perpendicular
($86^{\circ}$) to the line of sight.  In contrast, PSR B1706$-$44
has narrow $\gamma$-ray peak separation of 0.24 cycles,
and $\zeta = 53^{\circ}$
from its X-ray torus model.  \citet{ng08} summarize
the X-ray measurements and their implications for $\gamma$-ray
pulse profiles.

\begin{figure}[t]
\centerline{
\includegraphics[height=0.9\linewidth,clip=true]{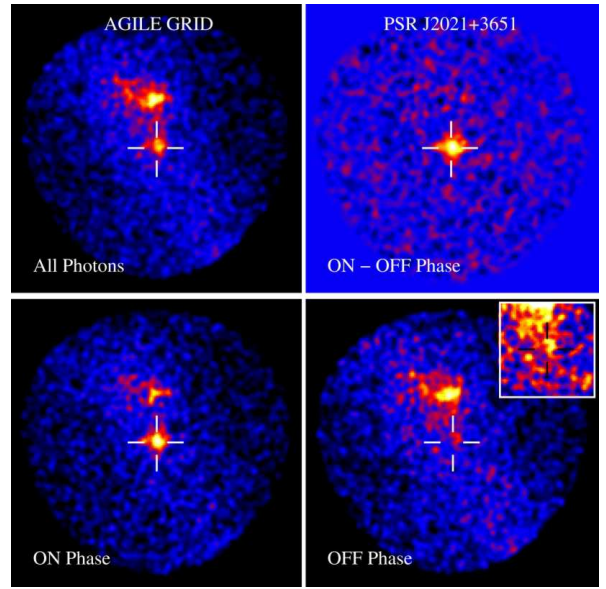}
}
\caption{Images of the \agile\ field centered on \psr\ ({\it cross})
in the 100--1500~MeV range.  Pulse-phase resolved images
were created as described in \S 3.2, and \psr\ was
isolated via difference imaging.  North is up and east is
to the left. The diameter of the field displayed is
$20^{\circ}$, and the pixel size is $0.125^{\circ}$.
Gaussian smoothing of $\sigma = 1.5$ pixels has been applied.
Each image is linearly scaled to its highest point (white), while
the inset stretches the scale to show the level of fluctuations
in the off-pulse image in the vicinity of the pulsar.
}
\label{fig:image}
\end{figure}

\begin{figure}[b]
\centerline{
\includegraphics[height=0.82\linewidth,angle=270.,clip=true]{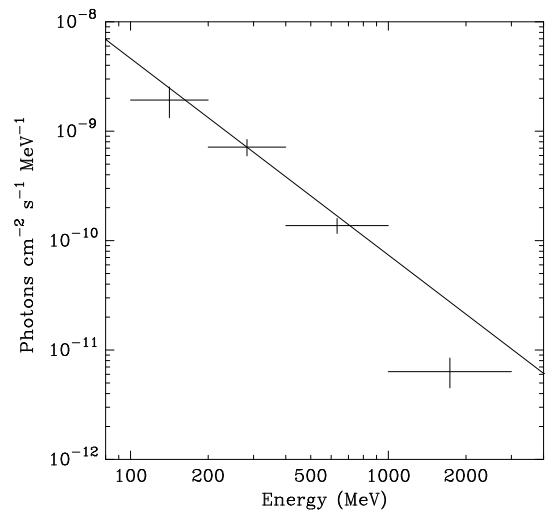}
}
\caption{Photon spectrum of \agl\ with power-law fit
of $\Gamma=1.86$ to the first three points.
}
\label{fig:spec}
\end{figure}

\subsection{Position, Flux, and Spectrum}

In order to separate the contribution of \psr\ from other possible
$\gamma$-ray sources within its PSF, we made
images corresponding to on-pulse and off-pulse
intervals of its light curve (Fig.~\ref{fig:image}).
The on-pulse image includes six phase bins on each of P1 and
P2, while the off-pulse
image comprises the remaining 28 bins.
A difference image was created by
subtracting the off-pulse image from the on-pulse
image, scaled by the number of phase bins.
As expected, differencing
removes all diffuse background and sources except for
\psr.  We measure the
position of the resulting source \agl\ as 
(J2000.0) R.A. = $20^{\rm h}20^{\rm m}30^{\rm s}$,
decl. = $+36^{\circ}53.7^{\prime}$,
which is only $0.12^{\circ}$ from the X-ray position,
well within the current \agile\ systematic uncertainty
of $0.25^{\circ}$ \citep{tav08}.
We performed a standard spectral analysis on the total source
photons including exposure correction and modelling of
the diffuse Galactic background.  The energy range
100$-$1000~MeV is fitted with a power law of photon index
$\Gamma = 1.86\pm 0.18$ and a flux of $(46\pm6)\times 10^{-8}$
photons cm$^{-2}$ s$^{-1}$, while a turndown is seen above 1.5~GeV
(Fig.~\ref{fig:spec}).

Using the counting statistics of the off-pulse image,
we estimate the level of fluctuation that would correspond
to a 3$\sigma$ point source at the location of \psr,
noting that there is no source present at this level.
Scaling from the pulsar flux, this corresponds to
upper limits of $1.0 \times 10^{-7}$ photons cm$^{-2}$ s$^{-1}$
for the off-pulse flux from \psr, and
$1.7 \times 10^{-7}$ photons cm$^{-2}$ s$^{-1}$ for a steady
point source within $1.2^{\circ}$ of \psr, 
such as the PWN \pwnb\ (of unknown spin parameters) or the blazar \blazar,
both located in the error circle of \egrb\ \citep{muk00,hal01}.
Since blazars are episodic $\gamma$-ray emitters, it cannot be ruled
out that \blazar\ was more active during the EGRET era.
However, we are not confident that \egrb\ was a source distinct from 
\egra.  The former is not included in the revised EGRET catalog
of \citet{cas08}, which instead lists only one source,
EGR J2019+3722.  The positions of \egra\ and EGR J2019+3722
are both inconsistent with \psr\ at their 99\% confidence
contours, while \agile\ sees one dominant source that is clearly
attributed to \psr\ by phase-resolved imaging.

\subsection{Gamma-Ray Efficiency}

Previous authors have noted that the $\gamma$-ray luminosity
of \egra\ is a large fraction $\eta_{\gamma}$ of the spin-down
luminosity of \psr\ if placed at the nominal DM
distance of 12~kpc using the \citet{cor02} electron-density
model of the Galaxy.
\cite{rob02} and \cite{hes04} calculated, assuming $1/4\pi$ beaming,
that $\eta_{\gamma} \equiv Fd^2/\dot E \approx 0.18\,d_{12}^2$,
where $F$ is the flux in the 100~MeV--10~GeV range.  Such
high efficiency deviates from the trend of other $\gamma$-ray
pulsars, which follow an approximate
$\eta_{\gamma} \propto \dot E^{-1/2}$ relation \citep{tho04}.
Pulsars of similar $\dot E$ to \psr\ have
$\eta_{\gamma} \sim 10^{-3}-10^{-2}$.
This contrast casts some suspicion on the accuracy of the
DM distance, but in addition, consideration of the X-ray
properties of \psr, including the thermal spectrum of
the neutron star and the size of its PWN torus,
led \citet{van08} to favor a distance as small as $3-4$ kpc.

Our results do not bear directly on the distance question,
while the inferred $\gamma$-ray efficiency of \psr\ remains
exceptional.
The \agile\ flux above 100~MeV is
$\approx 2.4 \times 10^{-10}$ ergs cm$^{-2}$ s$^{-1}$
yielding $\eta_{\gamma} \approx 0.10\,d_{12}^2$,
somewhat smaller than previous estimates 
because of the cutoff above 1.5~GeV.
In the models of \citet{rom95},
the beaming of \psr\ may be more severe than
in other $\gamma$-ray pulsars if its nearly
edge-on X-ray torus implies a large
magnetic inclination angle $\alpha$.

The TeV source \mgro\ \citep{abd07a,abd07b} 
of diameter $1.1^{\circ}\pm 0.5^{\circ}$
centered at $(\ell,b)=(75.0^{\circ},0.2^{\circ})$
may be an extension of \pwna, similar to many other
TeV PWNe that are spatially offset from their
smaller X-ray counterparts.  The flux of \mgro,
$\approx 5.6 \times 10^{-12}$ ergs cm$^{-2}$ s$^{-1}$,
represents $\approx 0.03\,d_{12}^2\,\dot E$ of \psr,
similar to other TeV PWNe \citep{gal08}.

\section{Conclusions}

After the demise of \cgro, several candidate
pulsar identifications for EGRET sources could not be confirmed
until it became possible again to obtain contemporaneous $\gamma$-ray
and radio timing.  Using the extensive exposure of \agile\
on the Cygnus region,
we detect $\gamma$-rays in the 100--1500~MeV
range from the first post-EGRET pulsar, \psr=\agl.
Its pulse shape is similar to other
$\gamma$-ray pulsars, with two sharp peaks separated by
slightly less than 0.5 cycles. This
morphology, as well as the lag between the
radio and first $\gamma$-ray peak, are in accord with simulations of
outer-gap accelerators viewed at large angles from the pulsar rotation axis,
which may help to explain its high $\gamma$-ray efficiency.  \psr\ is
the only $\gamma$-ray point source detected in the region of 2CG~075+00.
Timing noise of young pulsars such as \psr\ and other candidate
$\gamma$-ray pulsars demands frequent radio observations to
be able to discover and resolve the finest structure
in $\gamma$-ray light curves that will be accumulated
over months to years.  Independent discovery of pulsed
$\gamma$-rays from \psr\ by the {\it Fermi Gamma-ray Space Telescope}
will also be reported shortly.

\acknowledgements

We thank ASI for making these data available under the
Guest Observer Program of the \agile\ mission.
The GBT is operated by the National Radio Astronomy Observatory,
a facility of the National Science Foundation operated under
cooperative agreement by Associated Universities, Inc.
The referee David Thompson suggested
several improvements to this Letter.

\end{document}